\documentclass[10pt,conference]{IEEEtran}

\usepackage{float}
\usepackage[T1]{fontenc}
\usepackage{cite}
\usepackage{amsmath,amssymb,amsfonts}
\usepackage{algorithmic}
\usepackage{graphicx}
\usepackage{textcomp}
\usepackage{xcolor}
\usepackage[hyphens]{url}
\usepackage{fancyhdr}
\usepackage{hyperref}

\usepackage{tikz}
\usepackage{amsmath}
\usepackage{amssymb} 
\usepackage[table]{xcolor} 
\usepackage{array}
\newcolumntype{P}[1]{>{\raggedright\arraybackslash\begingroup\color{black}}p{#1}<{\endgroup}}

\usepackage{subcaption}

\usepackage[pro]{fontawesome5}
\usepackage{multirow}
\usepackage{amsmath}
\usepackage{url}
\usepackage{xcolor}
\usepackage[dvipsnames]{xcolor}
\usepackage{listings}
\usepackage{alloy}

\makeatother

\definecolor{gray}{rgb}{0.5,0.5,0.5}
\definecolor{bluekey}{rgb}{0.2,0.2,0.7}
\definecolor{greenstring}{rgb}{0,0.5,0}
\definecolor{orange}{rgb}{0.8,0.4,0}

\definecolor{codegreen}{rgb}{0,0.6,0}
\definecolor{codegray}{rgb}{0.5,0.5,0.5}
\definecolor{codepurple}{rgb}{0.58,0,0.82}
\definecolor{backcolour}{rgb}{0.95,0.95,0.92}
\title{Supporting Secured Integration of Microarchitectural Defenses }%

\author{
\IEEEauthorblockN{
Kartik Ramkrishnan,
Stephen McCamant, 
Antonia Zhai, 
Pen-Chung Yew
}
\IEEEauthorblockA{
Department of Computer Science and Engineering \\
University of Minnesota,  Minneapolis, MN, USA\\
\{ramkr004, mccamant, zhai, yew\}@umn.edu
}
}

\begin{document}

\maketitle

\pagestyle{plain}


\begin{abstract}
There has been a plethora of microarchitectural-level attacks
leading to many proposed countermeasures.  
This has created an unexpected
and unaddressed security issue where na\"{i}ve integration of
those defenses can potentially lead to 
security vulnerabilities.
This occurs when one defense  changes an aspect of a microarchitecture that is crucial for the security of another defense.
We refer to this problem as a \emph{microarchitectural defense
assumption violation} (MDAV).

We propose a two-step methodology  to screen for potential MDAVs in the early-stage of integration. The first step is to design and integrate a composed model, guided by bounded model checking of security properties.
The second step is to implement the model concretely on a simulator and to evaluate with simulated attacks.

As a contribution supporting the first step, we propose an event-based modeling framework, called \emph{Maestro}, for testing and evaluating microarchitectural models with integrated defenses. 
{\color{black} In our evaluation, Maestro reveals MDAVs (8), supports compact expression ($\approx$ 15x Alloy LoC ratio), enables semantic composability and eliminates performance degradations ($>$100x).}

As a contribution supporting the second step, we use an event-based simulator (GEM5) for investigating integrated microarchitectural defenses. 
We show that a covert channel attack is possible on
a na\"{i}vely integrated implementation of some state-of-the-art defenses,  
and a repaired implementation using our integration methodology is resilient to the attack.

\end{abstract}

\section{Introduction}
\label{section:introduction}

There is a large and growing variety of microarchitectural attacks, especially timing attacks~\cite{gruss2016flush+,yarom2014flush+,zhang2024timing,zaheri2025contention,o2024pixel,gruss2015cache,liu2015last,saileshwar2021streamline,zhao2024last,maar2025kernelsnitch} (see \S\ref{section:related_work}). In these attacks, it is possible for an attacker to spy on a victim's program by observing its microarchitectural side-effects on core states~\cite{kocher2019spectre,wikner2025breaking,wiebing2024inspectre,li2024indirector,chowdhuryy2024powspectre} and/or cache states~\cite{gruss2016flush+,morgan2025slice+,rueggebranch,rauscher2025systematic,luo2023autocat}. 
Attackers can also use them to establish covert channels and transmit secret information by bypassing architectural protections against such illegal communication.
To counter these microarchitectural attacks, a large number of defenses have been proposed~\cite{yan2018invisispec,saileshwarbespoke,ainsworth2021ghostminion,kessous2024prune+,gaudin2024fine,yin2025vericache,kvalsvik2023doppelganger} (see \S\ref{section:related_work}). 

Defenses are typically designed to counter a particular class of attacks, such as speculative attacks~\cite{kocher2019spectre}, cache hit-based attacks~\cite{yarom2014flush+} or cache-miss based attacks~\cite{gruss2015cache}. These defenses modify the core and/or cache design at the microarchitectural level so that covert channels and side channels are mitigated.

{\color{black} In the early-stage of design,
two important approaches for checking security properties are \textit{model checking}~\cite{alloy_docs,rosette} and \textit{attack simulation}~\cite{lowe2020gem5,yan2018invisispec,ainsworth2021ghostminion,aimoniotis2023recon}. Model checking enables exhaustive checks on abstract designs while finding security bugs in a few seconds or minutes~\cite{yang2023pensieve,hsiao2024rtl2mmupath}. It is a useful tool for finding property issues within a limited scale (e.g., thousands of bits), most often when a design is expressed abstractly with the most important features~\cite{yang2023pensieve}. Microarchitecture simulation enables us to test specific attacks in minutes or hours, on significantly more realistic designs, but it reverts to non-exhaustive testing~\cite{yan2018invisispec,ainsworth2021ghostminion}. 
Due to many non-overlapping strengths, these techniques are good for tandem usage in screening out insecure designs so that we can save on later-stage checking costs~\cite{tan2025rtl}.

}

However, \textcolor{black}{despite all of these advancements in early-stage secure design, a remaining important weakness of} the current practice is to design each defense in isolation without considering its impact on other defenses against different classes of attacks.

In reality, an attacker can choose which attack to implement and when to launch it. Therefore, we need to integrate ALL those defenses into the microarchitecture. 
When individual defenses are designed with only their particular classes of attacks in mind, it is likely that the assumptions and the specifications they used in their design may go against each other's.
It is forseeable that this could lead to what we call ``\textit{microarchitectural defense assumption violations} (MDAVs)'' and lead to new and unexpected vulnerability that can be exploited by attackers.
Hence, our key research questions for this paper are as follows.

\emph{\textbf{Can individual defenses against different classes of attacks cause unexpected security issues when they are integrated into a microarchitecture; and how can we check and identify such a possibility?}}

{\color{black} {\color{black} We propose an intuitive two-step methodology to answer the above research question.} 

The first step 
is to create a multi-defense \emph{model} 
that allows a designer to check the integrated defense for MDAVs.
}
If the model has an MDAV issue, the designer can propose a fix to the model. Then, the fixed model can be re-evaluated. {\color{black} The model checker can exhaustively check a small part of a system, but how that part fits in with the rest of the system can't be evaluated by the model checker.}
{\color{black} Hence, in the second step, the designer implements the design on a microarchitectural simulator, evaluates one or more attacks that target the MDAV and checks that the implementation of the fixed models is resilient to the attacks.

}

{ \color{black}
For the first step, the  designer builds models that represent individual defenses and then can put these defenses together to construct multi-defense models. The first  infrastructure challenge  is that integration of these models (to support multi-defense properties)  with na\"{i}ve patches can result in merge errors and build errors. Eliminating these errors from the integration workflow using semantic composition enables focus on the underlying MDAVs, especially when the designer wants to evaluate many possible combinations. 
}

{\color{black} Another key infrastructure challenge is that multi-cycle delays between events incurs  cost blowup for model checking.  
This is because current model checking frameworks~\cite{yang2023pensieve} lack  support for  event triggering with multi-cycle delay between steps. Instead, they express timing with cycle-by-cycle state changes~\cite{yang2023pensieve,hsiao2024rtl2mmupath}. As a consequence, expressing a large number of cycles, even if there are only a few events, blows up the runtime. This makes it impractical to represent and compose major defenses (see \S\ref{section:miscellaneous_evaluation}). 
}

{\color{black} To address these challenges, we propose an umbrella security term, \emph{\textbf{Maestro}}. It refers to a modeling framework
that natively supports \textbf{multi-cycle delays} between steps for eliminating blowups. It also refers to a composable transform strategy for supporting \textbf{semantic composition}, precluding the possibility of merge and build errors during composition. 

We implement Maestro by designing  domain-specific languages (DSLs)  for model specification (see \S\ref{subsection:am_dsl}) and semantic composition (see \S\ref{subsub:integra}).} We leverage Alloy~\cite{vakili2012temporal}, a standard modeling tool based on relational and linear temporal logic, as the backend model checker. Alloy itself uses a SAT-solver or SMT-solver backend to search the model space for violations of assertions (in our context, the assertions are security properties). { \color{black}
Using Maestro's semantic composition, the tool reveals  eight important MDAVs (see \S\ref{sub:generality}) and we show that in some cases, proper integration can avoid MDAVs. 

{\color{black} 
Even though Maestro cannot   model all the features in the model checking step (e.g., security-oriented updates to the hardware from system software), Maestro's results can guide a more concrete evaluation. 
As the  second step,
our methodology simulates a more concrete version of the defense on a simulator, namely, GEM5.
The model checker's results help with the implementation in the second step by either giving confidence in a secure design or by giving a counterexample as the basis for implementing an attack. Having the two steps work in tandem enables a relatively simple/abstract model to be effective.
}

{\color{black} 
With an implementation, a key challenge for a designer 
 is to realize a worst-case attack scenario to strain the implemented defense. Our methodology enables a designer to use the counterexample from the first step as inspiration. They can amplify the secret-dependent timing difference from the counterexample, in that particular  implementation. For instance, if the counterexample indicates that the timing difference depends on the duration of a certain speculative window, the challenge is to amplify it with a real sequence of (e.g., \texttt{x86\_64})   instructions. 
In \S\ref{section:attack_formulation},
 we demonstrate a scenario where the defense implementation  is resilient to a timing-amplified attack. 
 In \S\ref{section:related_work}, we note two important classes of side-channels that are re-opened due to MDAVs.
We also present an example in which considering how to build a defense on a simulator reveals a reason that the defense is impractical.
}

}

Overall, we have three major contributions in this work.
\begin{itemize}
     \item
     We identify the concept of \emph{ microarchitectural defenses assumption violations} (MDAVs) as an important issue for system designers working on secured microarchitectures. A two-step methodology is proposed to screen for MDAVs.
     \item
     For the first step of our methodology, we propose a systematic modeling framework, \emph{Maestro}, for investigating MDAVs. Maestro enables both cycle-based and event-based modeling that supports both detail and abstraction. { \color{black} In our evaluation, Maestro seamlessly enables semantic composition, enables $\approx$ 15x Alloy lines-of-code (LoC) ratio, discovers eight MDAVs and eliminates 100x performance degradations.} 
     \item
     For the second step of our methodology, we carry out a case study 
     of an integrated defense on the GEM5 simulator to test a  secured microarchitecture, which has resolved its MDAVs, against the original attacks and a new attack enabled by a {\color{black} timing-amplified instruction sequence}.

     \end{itemize}

To the best of our knowledge, this work  is the \emph{first}  systematic study of the integration of multiple microarchitectural defenses, thus enabling us to identify MDAVs. 
The rest of the paper is organized in the following manner.

\S\ref{section:background} provides background about different side-channel attacks and covert-channel attacks that are of interest to our integration study.  For the first-step of our two-step screening methodology,  \S\ref{section:framework} introduces the generic modeling framework, Maestro, for evaluating early-stage, event-based hardware models with integrated defenses. 
\S\ref{section:composition_aid} discusses a workflow using Maestro for effectuating defense integration. 
\S\ref{section:torc+dsrc} and 
\S\ref{section:miscellaneous_evaluation} discuss examples of integration using the workflow.

In \S\ref{section:attack_formulation}, we discuss the GEM5 implementation of \emph{start-with-S MESI} (SS-MESI) and \emph{Delay Speculative changes on Remote Miss} (DSRM) defenses, which are based on fixed models from the integration workflow. \S\ref{section:attack_formulation} and \S\ref{subsection:attack_experiments} implement a newly formulated covert channel attack on them and demonstrate their resilience. 
{\color{black} \S\ref{section:miscellaneous_evaluation} evaluates Maestro.}
\S\ref{section:related_work} discusses related work and miscellaneous issues. \S\ref{section:conclusion}  concludes the paper.

\section{Background}
\label{section:background}
We provide some background information for major attacks and defenses that are relevant to our problem of defense integration in this section. 
Spectre covert channels use speculative side-effects~\cite{kocher2019spectre,koruyeh2018spectre} in the cache~\cite{kocher2019spectre,ragab2024ghostrace} and other parts of the microarchitecture~\cite{koruyeh2018spectre,kocher2019spectre}. Major side-channel attacks include hit-based cache side-channels~\cite{yarom2014flush+,saileshwar2021streamline} and eviction-based cache side-channels~\cite{gruss2015cache,kessous2024prune+}.

\textbf{TORC Defense. } An important class of defense strategy is the TORC (\emph{Timing Obfuscation of Remote Cache lines}) strategy~\cite{ramkrishnan2020first,ojha2021timecache,yan2019secdir}, which obfuscates the cache hit time to make it appear to be the same as the cache miss time. This eliminates cache hit-based attacks.

\textbf{DSRC Defense. } 
Another important class of defenses are the speculative delay defenses~\cite{yan2018invisispec,ainsworth2021ghostminion}. This class of defenses eliminates speculative state changes in microarchitectural states, including the cache state. An important delay-based defensive strategy that protects coherence state from speculative leakage, is the DSRC (\emph{Delaying Speculative changes to Remote Cache lines})
strategy~\cite{ainsworth2021ghostminion}. In the DSRC approach, any load that changes vulnerable coherence state during cache hit (such as E/M state changed to S state), is disallowed from doing the change. Instead, a rejection signal is sent to the core. The core verifies that the \texttt{load} instruction is on the right path before re-issuing it to the  cache. Subsequently, a cache hit occurs and the rest of the load executes similarly to a baseline insecure processor.

\textbf{Integration of Multiple Defenses.} 
We may want to enable multiple defenses such as \emph{Speculative in-core Delay with Declassification} (SIDD)~\cite{aimoniotis2023recon, choudhary2021speculative}, \emph{Delay-on-Miss} (DoM)~\cite{ainsworth2021ghostminion,sakalis2019efficient},  \textit{Isolation}~\cite{ramkrishnan2024non,yan2019secdir} together,  
{\color{black} \textit{Speculative In-core delay with Data Obliviousness} SIDO~\cite{yu2020speculative} and \textit{Secure DRAM Refresh}  (SDR)~\cite{bostanci2025understanding}. SIDD enables declassification of speculatively accessed data based on whether non-speculative accesses already leaked them. SIDO enables speculative execution to occur while enabling operations to be independent of speculatively accessed data. \emph{Isolation}  partitions the caches to function independently of each other~\cite{kiriansky2018dawg}). \emph{Coherent Isolation} (CI) maintains data coherence between partitions~\cite{ramkrishnan2024non}). SDR  mitigates Rowhammer~\cite{mutlu2019rowhammer} attacks, which can cause unauthorized DRAM bit flips.
}

\section{Modeling Framework and Methodology}
\label{section:framework}

The key goal for the first part of our two-step methodology is to create a disciplined modeling framework.
In this section, we first present a high-level framework for studying the integration of defenses in \S\ref{subsection:framework}. 
 We then propose an implementation of the workflow using event-based modeling in Alloy Analyzer~\cite{vakili2012temporal} (a relational and temporal modeling tool) in \S\ref{subsection:am_dsl}.

\subsection{The Maestro Modeling Framework}

\label{subsection:framework}
Maestro is an abstract event-driven modeling framework in the context of microarchitectural-level security. It frames events as part of a step-wise execution that can trigger state changes and updates. 

\textcolor{black}{ Maestro facilitates the early-study of defense integration without requiring the full specification of  control/data paths and control/data logic. This is often the detail level at which defense designs are specified~\cite{ainsworth2021ghostminion,aimoniotis2023recon,saileshwarbespoke}. Maestro  enables the integration of defenses by allowing a defense model to add events to, or modify events of, a baseline model (see \S\ref{subsub:event_transforms}).}

\textbf{Machine State. } The machine state is a high-level abstraction of storage bits in a machine. It abstractly represents the data stored in registers, memory, buffers or other memory components in a processor. We represent the hardware containing the machine state as $\mathcal{H}$. Within $\mathcal{H}$, there are M bits [$B_1, B_2 ... B_M$].
Each $B_1$, ... , $B_M$ has a value of zero or one. 

\textbf{Stepwise Execution and Time. } An \emph{execution} consists of multiple steps with a possible state transition in each step. This represents a rich range of processor execution. We can represent the machine state in any step as $V^{x}$ where $x \in [0,N-1]$, and $N \in \mathbb{N}$ is the total number of steps. During each execution step $V^{x}$, the machine state contents can change to something different from the previous step. Hence, in the transition sequence $V^{0} \rightarrow V^{1} \rightarrow... V^{x}... \rightarrow V^{N-1}$, the corresponding machine state in any step $x$ is $[B_{1}^{x}, B_{2}^{x} ... B_{M}^{x} ]$. The initial state of the system is termed as $V_0$ and the state at step $x$ is termed as $V_x$. 

\emph{Compressed Sequence.} In a compressed sequence, each step is associated with a \emph{time}. Time is a non-negative integer. A concrete example of what time could represent is a clock in a synchronous circuit.
Time can increase by one or more cycles in each step, and always increases at least by one cycle. 

Using this extension, we can now represent a time sequence of state changes as $V^{0}_{t_0} \rightarrow V^{1}_{t_1} \rightarrow... V^{x}_{t_x}... \rightarrow V^{N-1}_{t_{N-1}}$. The step counts always increase by one, but the time can increase by more than one cycle. 

\textbf{Events.} 
A Maestro model has a list of \emph{event specifications}.
Each event specification in the list $[E(\alpha,\beta,\gamma,\delta,\epsilon)]$ (i.e., $\mathcal{E}$ in short), consists of an event specification name (${E}$), \emph{carried data} ($\alpha$), a conditional sequence of \emph{event triggers} ($\beta$), a conditional sequence of \emph{state transitions} ($\gamma$), a \emph{time-delay} ($\delta$) and a \emph{PresentAtStart} flag ($\epsilon$).

Any event instance $e$($\alpha, \beta, \gamma, \delta, \epsilon$) of an event specification contains the following data:  
$\alpha$ contains the fields [$d_1$, $d_2$ ... $d_r$], each of which has a bit value 0 or 1.  $\beta$ contains conditions for triggering an event [$c_1$: $E_1$, $c_2$: $E_2$ ... $c_p$: $E_p$] in which
$c_i$  is a Boolean function based on the current machine state and on $\alpha$ and $E_i$ is an event specification from the list.

$\gamma$ contains the state transitions and the conditions that trigger them. It is represented as a list [$cc_1$: $st_1 \leftarrow NV_1$, $cc_2$: $st_2 \leftarrow NV_2$, ..., $cc_q$: $st_q \leftarrow NV_q$].  
The $\delta$ field shows
how much time it should pass before it is ready to trigger its child events and state transitions. 
This means, if an \emph{event} appears in a particular \emph{step} $x$ at the time $t_x$, then its child event only becomes `active' in the step $y>x$ at the time $t_y\ge t_x+d$, i.e., a child event can be triggered in a later step. The \texttt{PresentAtStart} flag $\epsilon$ indicates whether the event instance is present at step 0. It can take the value 0 or 1.

\textbf{Event Sequence.}
An execution is an \emph{event sequence} ($V^{0}_{t_0},[e]^{0}_{t_0}) \rightarrow (V^{1}_{t_1},[e]^{1}_{t_1}) \rightarrow... V^{x}_{t_x},[e]^{x}_{t_x}... \rightarrow V^{N-1}_{t_{N-1}},[e]^{N-1}_{t_{N-1}}$.

The $[e]^{x}_{t_x}$ represents all the event instances during a particular step x.

In the first step ($x=0$), there will be particular state values and event instances specified as the initial condition.
Within any step $x >= 0$, we check all the triggering conditions, i.e., the Boolean functions $c_i$ in its $\beta$. If any condition $c_i$ is satisfied in the step ($x$), then an instance of the corresponding event $E_i$ is triggered in the next step ($x+1$). The timing difference between the next step and the current step, i.e., ($t_{x+1}-t_{x}$), is set to be one in this case.
If we check all events in the preceding step and find that there are some events that have not reached their delay thresholds. These events will continue onto this step.
Finally, if there is an event in the previous step that has reached its trigger threshold but none of its $c_i$ conditions is active, in this case, this event will also be present as is in the next step. The timing will be adjusted to be one greater in this condition because we don't know when it will eventually be triggered.

\begin{figure}
  \includegraphics[trim=0 0 0 0, clip, width=\linewidth]{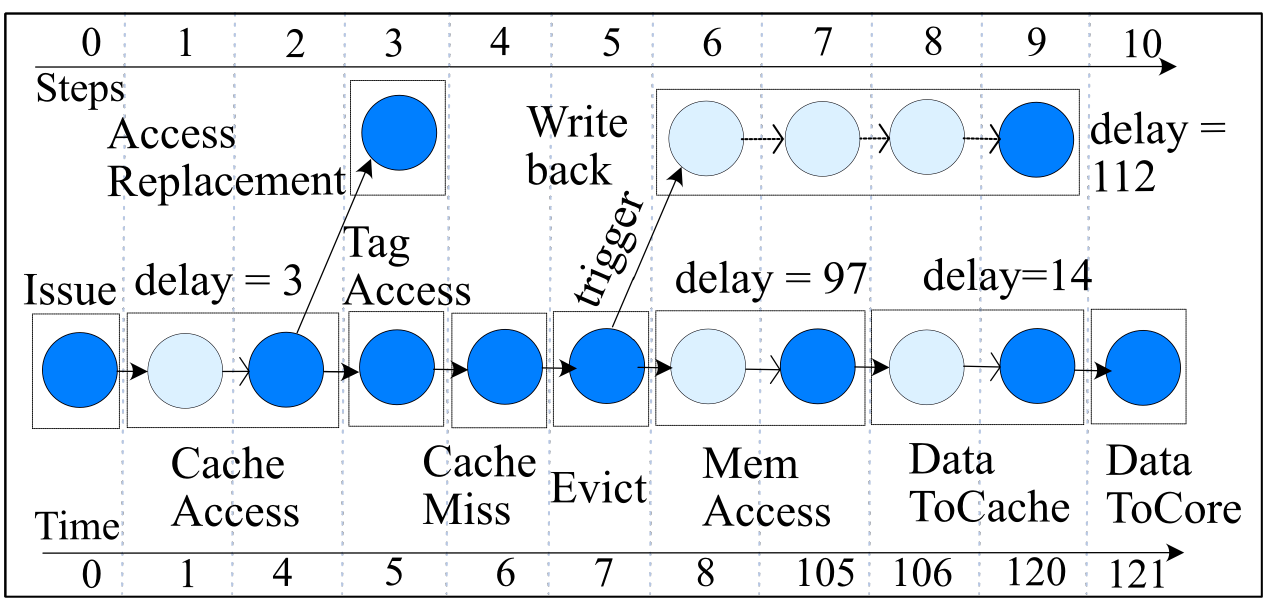}
  \\[-3ex]
  \caption{\textmd{An example illustrating key concepts in the construction of an event tree. A box represents an event instance. In a given step (column), an event exists in an active state (dark blue circle) or a pending state (light blue circle). An event always ends in an active state. 
  The event tree represents the event sequence of a cache miss as explained in  \S\ref{subsection:framework}}.}
  \label{fig:model_example}
\end{figure}

\textcolor{black}{
\textbf{Event Tree.}
Based on the above constraint specification, an event sequence actually forms an \emph{event tree} according to the events'  triggering conditions and timing.
An input configuration gives a value to initial machine state and state of initial events.
Different input configurations (i.e., different initial conditions) can form a different event tree.
We use the following example to elucidate the concept of an event tree.}

\textcolor{black}{
 In Figure~\ref{fig:model_example}, we start with an \texttt{Issue} event in the initial step $s^{0}_{0}$. The superscript contains the step number and the subscript contains the time value. The \texttt{Issue} event triggers a \texttt{CacheAccess} event in $s_1^{1}$. Assuming the specified delay $\delta$ is 3, the \texttt{CacheAccess} event remains pending (shown as a light blue node) until step $s^{2}_{4}$. The \texttt{CacheAccess} event then triggers an \texttt{AccessReplacement} event and a \texttt{TagAccess} event at the same time in $s^{3}_{5}$. Following the \texttt{TagAccess} event, there is a \texttt{CacheMiss} event in $s^{4}_{6}$ due to tag mismatch in the \texttt{TagAccess} event, otherwise there would have been a cache hit event, leading to a different event tree. The \texttt{CacheMiss} event triggers an \texttt{Evict} event in $s^{5}_{7}$. This eviction triggers a \texttt{Writeback} event because the evicted cache line is dirty and a \texttt{MemAccess} event for the missed cache line. In the $s^{6}_{8}$ these two events appear. The \texttt{MemAccess} event stays until the next step $s^{7}_{105}$ to satisfy its delay requirement of 97 cycles for its memory access latency. It then triggers a \texttt{DataToCache} event. This event is triggered in the next step $s^{8}_{106}$ and is pending for another 14 time units to send the data to the cache in the next step $s^{9}_{120}$. Simultaneously, in $s^{9}_{120}$, the \texttt{Writeback} event has satisified its delay requirement of 112. In the last step, $s^{10}_{121}$, the \texttt{DataToCache} event triggers \texttt{DataToCore}.   }

\textbf{Security Property. }  
At each step of  execution, we can  enforce a security property $\mathcal{S}(Sequence)$ that is a true or false function of the event sequence so far. For the security property to be satisfied, the true condition should always be true. If it is false at any step, it means that the property has been violated.

\textcolor{black} {A key security property is the \emph{non-interference} property. It means that an attacker would not be able to observe the influence of secret-dependent state. Non-interference is formalized by considering two different executions of a system differing in secret-dependent state, and requiring  the observations of an attacker on the two executions to be the same. It can be represented in a Maestro model as two event trees (representing two identical machines), each accessing a copy of an input configuration  which is the same except for  different initial secret-dependent values.}



\begin{lstlisting}[caption={\textmd{An example using the Maestro DSL represents a simple circuit that updates a 5-bit counter every step  \S\ref{subsection:am_dsl}}.}, label={lst:alloy_maestro}, captionpos=b, float]
#Maestro DSL 5-Bit Counter Example
---
MachineState:
  - TypeSpec:
     - Counter: {entry: BV[5]}
  - InstanceSpec:
     - ctr1: Counter
Events: 
  - Name: "ClockEdgeEvent"
    CarriesData: "None"
    TriggersEvent: "Trigger ClockEdgeEvent{NONE}"
    StateChanges: "SC ctr1.entry <- ctr1.entry+1"
    TimingDelay: "0"
    PresentAtStart: "Yes"
Assertions:
   - Name: "Incrementing_Counter"
     Assert: "ALWAYS ctr1.entry'=ctr1.entry+1"
InitialState:
   -  Constraint1: "ctr1.entry = 0"
MaxSteps: 33 
IntWidth: 7 
\end{lstlisting}

\subsection{A DSL for Implementing Maestro}
\label{subsection:am_dsl}
Maestro is a \emph{domain-specific language}  (DSL) that builds on a YAML-based format~\cite{yml_docs} to create a microarchitectural specification for the Maestro framework.
There are several sections in its specification: 
\emph{MachineState, Events, Assertions, InitialState, MaxSteps,} and \emph{IntWidth}.
We present each section in the followings using a 5-bit counter as an example (see \textcolor{black}{Listing~\ref{lst:alloy_maestro}}).

\textbf{Machine State ($\mathcal{H}$). } The first section in the specification is the description of the machine state.

There are two parts in the section. The first part is the \texttt{TypeSpec} which lists all the different types of hardware components in the system.
The second part is the \texttt{InstanceSpec} which lists and names each instance of the components in the system. For example, we can have a data cache and an  instruction cache, which are both instances of the same cache type. 
Together, the \texttt{InstanceSpec} and \texttt{TypeSpec} represent the machine state. 

 In \textcolor{black}{Listing~\ref{lst:alloy_maestro}}, there is one system component whose type is \texttt{counter}.
 It has one instance named \texttt{ctr1}.

 The size of its machine state is specified in the \texttt{entry} field, i.e., \texttt{ctr1.entry}, which is \texttt{BV[5]} in this example (a 5-bit bitvector). 

 \textbf{Events ($\mathcal{E}$).} This section provides different event specifications of the system. Each event has a \texttt{Name} field that provides a name for the event specification.
\texttt{CarriesData}, \texttt{TriggersEvent}, \texttt{StateChanges}, \texttt{TimingDelay}  and \texttt{PresentAtStart} correspond to $\alpha$, $\beta$, $\gamma$, $\delta$ and $\epsilon$.

In  Listing~\ref{lst:alloy_maestro},
the name of the event for the counter is \texttt{ClockEdgeEvent}.
\texttt{CarriesData} contains a list of data fields in the event. 

As there is no data carried by the \texttt{ClockEdgeEvent}, it is marked as \texttt{None}. 
The \texttt{TriggersEvent} specification allows the event to conditionally or unconditionally trigger another event and to assign values to the data fields inside that event as needed. 
For the counter,
this field is \texttt{Trigger ClockEdgeEvent \{NONE\}}, which means that the \texttt{ClockEdgeEvent} is triggered unconditionally in the next step, and that there is no data assignment to that \texttt{ClockEdgeEvent}. 

 The \texttt{StateChanges} specification represents a sequence of conditional state updates. In Listing~\ref{lst:alloy_maestro}, the \texttt{StateChanges} specification is \texttt{SC ctr1.entry <- ctr1.entry+1}. This means, the counter value is incremented by one when a \texttt{ClockEdgeEvent} is triggered.  
 The \texttt{TimingDelay} field is a non-negative integer that indicates the number of additional time units before a triggered event can complete. The \texttt{PresentAtStart} field is a ``Yes'' or a ``No'' string that indicates whether an instance of the event specification is present at step zero. The complete grammar for $\alpha$, $\beta$ and $\gamma$ that is used to write the YAML specification and compile it to Alloy is provided in the online supplement~\cite{supplement}.  

\textbf{Assertions ($\mathcal{S}$). } 
The assertions specify security properties such as non-interference and constant-time requirements. Assertions are used to check the security property during every step of the execution (\texttt{ALWAYS}) or at the end of the execution (\texttt{FINALLY}).
In Listing~\ref{lst:alloy_maestro}, the assertion is \texttt{ALWAYS ctr1.entry'=ctr1.entry+1}. It means that the value of the counter entry in the next step is always one more than the counter entry in the current step.

\textbf{Initial State $[B_1^{0}, B_2^{0} ... B_M^{0}]$. } 
By default, the initial state  has no constraint. 
If there are any constraints on the initial state, they are specified in the \texttt{InitialState} field. 
In Listing~\ref{lst:alloy_maestro},  "\texttt{ctr1.entry = 0}" means that the initial state of the counter is zero. 

 \textbf{Max Steps (N). } The next field of our YAML specification is the maximum number of steps.
 This field bounds the search space of the model. It should be large enough to encompass the complete functionality that we are checking. 
 This can be set depending upon the context. In Listing~\ref{lst:alloy_maestro}, the value is 33, which covers the case where the 5-bit counter wraps around. 

 \textbf{Int Width. } This field determines the width of the \texttt{Int} field in the finally generated Alloy code. A higher \texttt{Int} field increases the range of timings and steps that we can test our model on while also increasing runtime. 
 In Listing~\ref{lst:alloy_maestro}, it needs to be at least 7 to incorporate 33 timing steps.

\begin{lstlisting} [language=Alloy, caption={ \textmd{An Alloy snippet demonstrating signatures and constraint initialization, generated from Maestro Listing~\ref{lst:alloy_maestro}.}},  label={lst:alloy_maestro_code_snippet}, captionpos=b, float]
-- Module and Signature Definitions
1 open bitvector as bv
2 sig ClockEdgeEvent { 
3 var status: Int, var appearance_time: Int, var delay: Int, var event_id: Int, var reason:Int, var parent_id: Int }
4 one sig ctr1_entry {var val:BitVec5}
5 one sig TimingRecord{var time: Int}
6 one sig StepRecord{var step: Int}
7 -- Timing and Steps
8 fact{always{
9  StepRecord.step < 32 => {
10    StepRecord.step' = add[StepRecord.step,1]
11    TimingRecord.time' > TimingRecord.time }}}
12 -- Initial State
13 fact{(one e: ClockEdgeEvent | e.status >= 1) and bitVecFromBits5[Zero, Zero, Zero, Zero, Zero, ctr1_entry.val] and StepRecord.step = 0 and TimingRecord.time = 0
14      all e:ClockEdgeEvent | e.status >= 1 => (e.appearance_time=0 and e.delay = 0)}
17 -- Range and uniqueness constraints
18 fact{always{
19  /////// Unique event ID constraint ///////
20  /////// Event ID, Parent ID range constraint ///////
21  /////// Status field range constraint ///////
22  /////// Event instance counts constraint ///////
23  /////// Tie bitvectors to machine state ///////
24 }}
\end{lstlisting}

\subsection{ Translating  Maestro's DSL to Alloy}
\label{subsection:am_dsl_conversion}

We present the implementation approach of the translation from Maestro to Alloy below. We subsequently discuss the key objects and relations  in \S III-C.1. 
Event sequences and assertions are discussed in \S III-C.2. Listings~\ref{lst:alloy_maestro_code_snippet} and \ref{lst:alloy_maestro_code_snippet_two} are an abridged version of Alloy generated from a Maestro specification. Comments are introduced with a `\texttt{--}'   or `\texttt{///////}'. `\texttt{///////}' at the end of a comment indicates that the code for that section is omitted due to space constraints.

 \textbf{Relations and Objects in Alloy. }  

 The key features of Alloy in this context are to set up  objects (i.e., atoms), relations between the objects and changes of those  relations over different steps.  
 \emph{Objects} are the most basic unit in Alloy.  Alloy defines a \emph{relation} between two (or more) sets of objects. All information about objects is represented with relations.  For example, the value of a counter is a relation between a counter object and an integer object, such as `5'.
Every object belongs to a \emph{signature}, similar to a class in Java/C++. For this example, the syntax (\texttt{sig counter \char`{val: Int\char`}}) defines an object class, where each object of the \texttt{counter} class  has a relation \texttt{val} to an integer object. 

\emph{C.1) Initialization and Definitions}

\textbf{Steps and Time. } 
Alloy has the concept of `step' but it does not have the notion of `time'. Also, it does not maintain a step count on its own. To represent these concepts, we define a \emph{step record} and a \emph{timing record}. The signatures are constrained to have one object of each class, and both step and timing records are initialized to zero in the first step. The step record  is constrained to increase by one   in every step. The timing record is constrained to increase in every step but the amount is not constrained.
The corresponding constraint declarations in Alloy (lines 10 and 11 of Listing~\ref{lst:alloy_maestro_code_snippet}) are \texttt{StepRecord.step'=add[StepRecord.step,1]} and \texttt{TimingRecord.time'>TimingRecord.time}.
The \texttt{\char`'} notation (\texttt{step\char`'} instead of \texttt{step}) on the left-hand side indicates that a relation is changing in the \emph{next} step, also known as a \emph{mutation} in Alloy. 

\textbf{Machine State.} 
We define machine state as Alloy objects using the \texttt{sig} keyword. The Maestro backend translates each entry in the Maestro machine state (from the \texttt{InstanceSpec} and the \texttt{TypeSpec}) to create an Alloy signature.
Listing~\ref{lst:alloy_maestro_code_snippet}, line 4, shows a representation of machine state in Maestro. In this case, the generated object signature for the Maestro machine state \texttt{ctr1.entry} is \texttt{one sig ctr1\_entry \char`{var val: BitVec5\char`}}. This means that there is an object called \texttt{ctr1\_entry} \textcolor{black}{and that it has a relation to a \texttt{BitVec5} object, called \texttt{val}}.
The \texttt{one} before the \texttt{sig}  keyword specifies that there is exactly one object for this signature. Absence of the \texttt{one} keyword means there could be any number of objects.

\textbf{Events. } Event specifications are defined by signatures.
Event instances are represented by objects of that signature.
Listing~\ref{lst:alloy_maestro_code_snippet}, lines 2--3,
shows the event signature for \texttt{ClockEdgeEvent}.
The \texttt{status} relation specifies where the object is in its lifecycle. It can be one of \emph{undeployed} (0), \emph{deployed-but-pending} (1) or \emph{deployed-and-active} (2).
The \texttt{appearance\_time} relation specifies the time at which the object instance was deployed. \texttt{delay} represents the timing delay before a triggered event can exit. \texttt{event\_id} is a unique integer assigned to an event object. \texttt{reason} is the index $i$ of the condition $c_i$ that causes an event to be triggered. \texttt{parent\_id} is the \texttt{event\_id} of the triggering event.

\textcolor{black}{\emph{Initial Constraints. }
The initial constraints (at the first step) on relations, specified using  \texttt{fact}, describe the initial state of the system.   Listing~\ref{lst:alloy_maestro_code_snippet}, line 13,  constrains the \texttt{ClockEdgeEvent} object's \emph{status} to be at least 1. It also constrains the counter object's value  to be zero (using a \texttt{bitVecFromBits5} predicate), and the step and time value to be zero. Line 14 constrains all deployed \texttt{ClockEdgeEvent}s  to have a zero appearance time and  delay.}

\textcolor{black} {\emph{Range and Uniqueness Constraints. } The range and uniqueness skeleton is shown from line 19-24 in Listing~\ref{lst:alloy_maestro_code_snippet}. The \emph{unique event ID} constraint specifies that the each event object has a different integer \texttt{event\_id}. The \emph{event ID range} constraint ensures that events with different objects with the same signature have different event IDs. The \emph{status range} constraints require the status field to have a value 0, 1 or 2. The \emph{event instance count} constraints specify the maximum number of objects for each event signature. 
The \emph{tie bitvectors} code constrains each machine state to reference a unique bitvector object. }
\emph{C.2) Event Sequences, State Updates and Assertions}

There are four parts in the event sequence construction, namely, \emph{deployment}, \emph{maintenance}, \emph{state updates}, and \emph{completion} (from line 4 to line 8 of Listing~\ref{lst:alloy_maestro_code_snippet_two}). 
Each of these different parts serves to enforce updates to relations  that together implement the Maestro specification.
The \emph{deployment} part enforces triggering of events, i.e., it ensures that there is an update of a \texttt{status} relation from 0 to 1, and it also enforces the data relations of the object. The \emph{maintenance} part specifies constraints that need to be true so that the \texttt{status} relation of an object is 1 (deployed-but-pending) or 2 (deployed-and-active) in the next step. If the delay does not expire in the next step, the \texttt{status} is 1, otherwise it is 2. The constraints in the \emph{state updates} part specify that the relations of the machine state objects track with event completion.
The \emph{completion} part constrains events to exit once they have completed their delays, i.e., the \texttt{status} relation updates from 2 to 0.

\textbf{Assertions. }
The \texttt{ALWAYS} assertion in Maestro is expressed in Alloy as an \texttt{always} block inside an \texttt{assert} statement. Examples using this pattern are shown from lines 10--17 of Listing~\ref{lst:alloy_maestro_code_snippet_two}.  
The first assertion (lines 10--12, \texttt{alwaysOneClock}) checks that the count (\texttt{\#}) of the \texttt{ClockEdgeEvent} objects that have a \texttt{status} field at least 1, is one. The second assertion (lines 13--15) checks that the bitvector \texttt{ctr1\_entry.val}, always increments by one every step (\texttt{addBitsToVec5} predicate). 
The third assertion  (lines 16--17) checks that the integer time value in the next step (\texttt{TimingRecord.time'}) is always one more than the time value in the current step. The \texttt{add} function does the increment-by-one and the \texttt{=} operator compares the integer time values.

\texttt{check} statements evaluate the assertions. For each \texttt{check} statement, Alloy generates a counterexample if the assertion fails. 

The \texttt{check} feature of Alloy identifies any MDAVs in the Maestro model.
The relation-mutation sequence of the counterexample is translated back to a Maestro event tree by a helper script.

\textbf{Max Steps and Int Width. } The maximum steps can be set within Alloy using the \texttt{step} keyword (33 for this example). 
The bitwidth, 7, of the \texttt{Int} objects is set as \texttt{7 Int} (see line 18 of Listing~\ref{lst:alloy_maestro_code_snippet_two}).

\begin{lstlisting} [language=Alloy, caption={ \textmd{ Event lifetime constraints and assertions, generated from a Maestro specification in Listing~\ref{lst:alloy_maestro}.}},  label={lst:alloy_maestro_code_snippet_two}, captionpos=b, float]
... Initialization ... 
1 -- Lifecycle constraints
2 fact{always{ 
3 StepRecord.step < 32 => { 
4  /////// DEPLOYMENT, MAINTENANCE, STATE UPDATES, COMPLETION    ///////
8 }}}
9 -- Assertions to check
10 assert alwaysOneClock {
11     always {StepRecord.step < 32 => (#{e:ClockEdgeEvent | e.status >= 1} = 1)}
12 }
13 assert alwaysIncrementCounter {
14      always { StepRecord.step < 32 => addBitsToVec5[One, Zero, Zero, Zero, Zero, ctr1_entry.val] }
15 }
16 assert alwaysIncrementTime {
      always { StepRecord.step < 32 => (TimingRecord.time' = add[TimingRecord.time, 1])}
17 }
18 run {} for 33 steps, 7 Int
19 check alwaysOneClock for 33 steps, 7 Int
20 check alwaysIncrementCounter for 33 steps, 7 Int
21 check alwaysIncrementTime for 33 steps, 7 Int
\end{lstlisting}

\section{Two-Level Model-Integration Workflow Using Semantic Composition}
\label{section:composition_aid}

\textcolor{black} {To constrain the factorial complexity into a practical workflow,
an intuitive strategy is to split the integration process into \emph{two} levels. In the first level, we have the different models that need to be integrated. In the second level,
we have an accumulated model, and a selected model that is the current focus of integration.}

Based on the above two-level strategy, we propose a model integration workflow. In this workflow, the designer starts with a selection of defenses that they wish to integrate in turn. To facilitate this integration, we propose a composable defense model definition strategy in terms of a baseline Maestro model (without defenses) and  the defenses  that they want to integrate (e.g., Spectre protections), expressed as \emph{transformation functions}. We call them \emph{transforms} in short (see \S\ref{subsection:patching_strategies}).

 The designer iteratively  integrates a new Maestro defense model from the selection with  previously integrated defenses and checks for the existence of an MDAV (expressed as a Maestro event sequence obtained from an Alloy \texttt{check} statement's counterexample). If there is an MDAV, the designer revises the defenses and integrates again. The iterations continue until there are no MDAVs or if there is no apparent solution~(see \S\ref{subsub:integra_workflow}). 
\subsection{Transforms for  Supporting Composable Models}
\label{subsection:patching_strategies}

\subsubsection{Event Transforms}
\label{subsub:event_transforms}

Event transforms can add events to the event specification list. They can also modify an existing event specification but only in a composable way. For example, they can add data fields to an event but they cannot remove data fields. Another example is that Maestro permits the designer to use either logical AND or logical OR transforms to an event's Boolean conditions, but not both. 

\subsubsection{Machine State, Assertions and Initial State}   
The machine state transform adds machine state to the Maestro model.  

The assertion and initial state transforms add assertions and initial state constraints into the assertion and initial state lists. Full details of our composable event transforms are in the online supplement~\cite{supplement}. 

\subsubsection{Model Composability}
The order of composing the defenses does not affect the composed model. For example, if we have defenses A and B added to a baseline model, the order of composition, baseline + A + B yields the same composed model as baseline + B + A. 

\subsection{Non-Interference Transform}
\label{subsub:non_interference}

A \emph{non-interference} property means that an attacker would not be able to observe the influence of secret-dependent state. Non-interference is formalized by considering two different executions of a system differing in secret-dependent state, and requiring  the observations of an attacker on the two executions to be the same.

For modeling a non-interference property in Maestro, we duplicate the machine state and event
specifications of the given model so that we effectively have two machines. In the initial step, the secret-dependent state 
is 
unconstrained so that it can be different between the two machines,  while the rest of the state is 
constrained to be the same in the two machines. For checking a non-interference property, an \texttt{ALWAYS} assertion checks that the observable machine states are always the same in both machines. See \S\ref{subsection:TORC} for an example. 
This transform is not for adding defenses, it is to check for MDAVs.

\begin{figure}[b]
  \includegraphics[trim=0 80 0 20, clip, width=\linewidth]{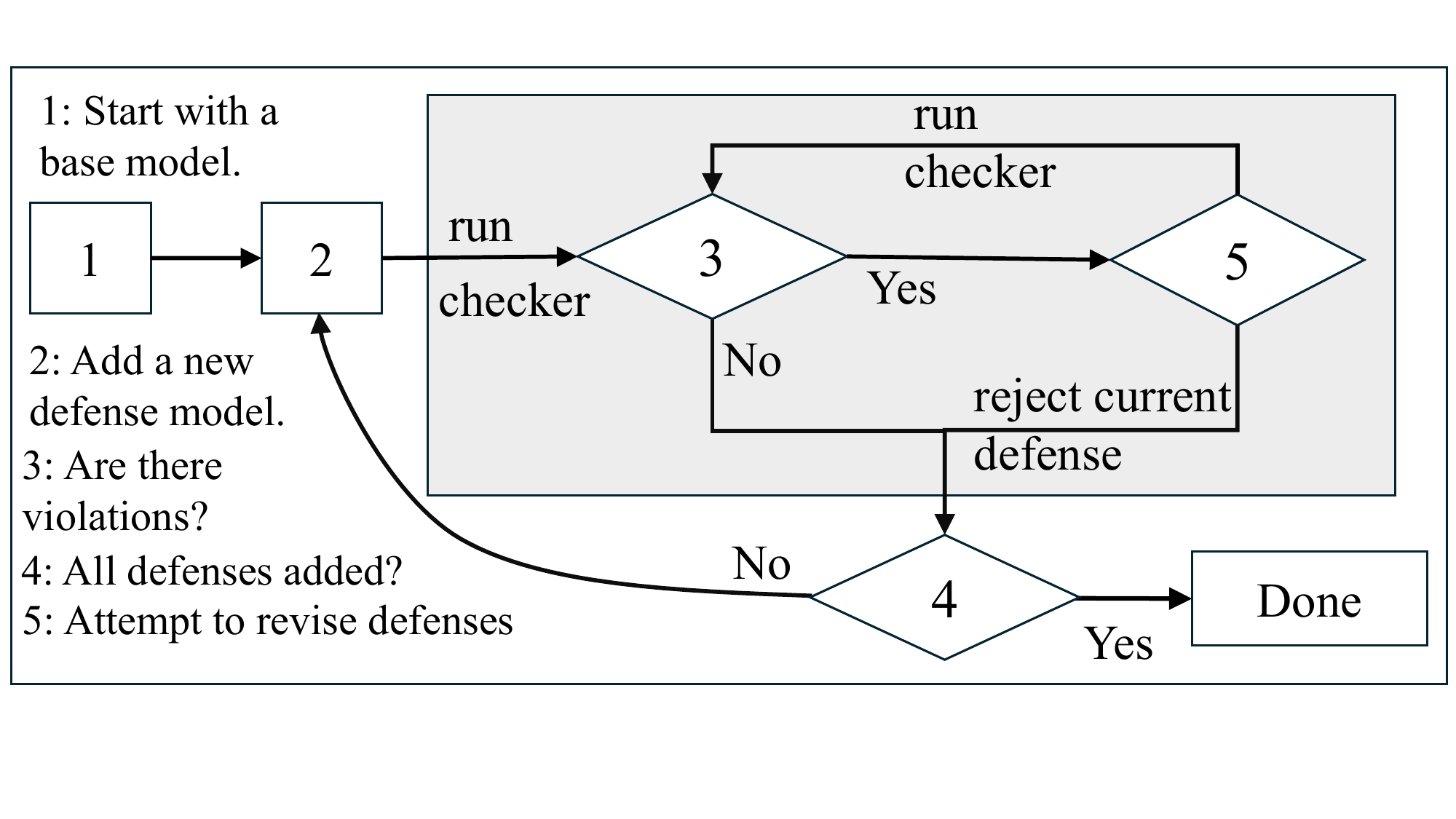}
  \\[-3ex]
  \caption{\textmd{The model integration workflow that we use to study the interaction between multiple defenses. We add defenses one at a time to build up a larger defense while allowing for the possibility to reject violation-causing defenses that we cannot revise (see \S\ref{subsection:framework})}.}
  \label{fig:framework}
\end{figure}

{\color{black} 
\subsection{Integra Transforms}
\label{subsub:integra}
To implement transforms with the desired compositional properties, we implement another DSL within Maestro, which we call \textit{Integra}. In Integra, there are 15 composable transformations~\cite{supplement}. 
Two or more Integra programs are composed by collecting transforms into one program. Four key transforms are used for a non-interference check. 

The first two effectively create a copy of a modeled system with similar states but potentially different state values. The third transform permits a secret-dependent difference in an initial value for a state and its counterpart in the other machine. The fourth transform checks that a particular state and its counterpart in the other machine always have the same value (non-interference).}

\subsection{Model Integration Workflow Using Maestro}
\label{subsub:integra_workflow}

The designer starts with two or more Maestro models, each representing a distinct defense for a common baseline model.

Figure~\ref{fig:framework} shows the defense integration workflow in a flowchart. 
In step (1), the workflow starts with a baseline model. In step (2) the designer chooses a defense model that has not yet been integrated. Then, this model is composed with the already-integrated model. 
In step (3), the resultant Maestro specification is then checked to see if it satisfies the security properties (including those that use non-interference transforms). The workflow then transitions to step (4) if there are no MDAVs. In step (4), the workflow either concludes or moves to step (2). If no more defense models are to be added, the workflow concludes. Otherwise, the next defense model is selected in step (2). On the other hand, if the composed model does not satisfy the required security properties, the designer revises the defenses in step (5), using  the provided counterexample, and then goes back to step (2).

In step (5), if the designer cannot think of revision, then the the designer can choose to move to step (4).

\section{Example: Maestro Integration}
\label{section:torc+dsrc}
In this section, we use the  TORC and DSRC defenses to illustrate the Maestro framework's integration workflow. This is an example of the first step of our two-step MDAV screening methodology.
We select TORC and DSRC for our study because they target different but important attacks, namely, Spectre attacks that use cache coherence state and cache flush based side-channels.  Unlike DSRC, TORC defenses eliminate cache-hit based attacks, which do not necessarily use speculative execution.
We model a simple processor with two cores as a baseline in Maestro that executes a load instruction which may or may not be speculative. 

The baseline model and the defense transforms are discussed in  \S\ref{subsection:TORC}. We  integrate the two models using our workflow in \S\ref{sub:secure_combination_torc_dsrc}.

\subsection{Baseline Model and Transforms}
\label{subsection:TORC}
\textbf{Baseline Model.} In the event sequence for the load instruction, we start with an \texttt{IssueEvent}, then a \texttt{CacheHitEvent} or a \texttt{CacheMissEvent}. If it is a \texttt{CacheHitEvent}, then a \texttt{CompletionEvent}  occurs next. Otherwise, there is a \texttt{MemoryAccessEvent}, followed by a \texttt{CacheUpdateEvent}  and then a \texttt{CompletionEvent}.

The relevant machine state to represent a cache line is a cache line address,  a presence/absence bit, a sharer bit vector and a coherence state bit exclusive/shared (or E/S). The E state is represented by bit value 0 and the S state by bit value 1. 
There is a completion bit, which is set when the load completes in the core. 
In the initial state, the completion bit is set to zero. If the presence bit is zero, then the other machine state bits in the cache line are zero. Also, the coherence bit is S if the sharer vector has more than one bit set.

The baseline model does not have any security properties that need to be enforced (by  assertions). However, the load instruction should function correctly. Specifically, the model requires
that the completion bit be set at the end of the execution. This is expressed as a \texttt{FINALLY} assertion.  

\textbf{Timing Obfuscation of Remote Cache Lines (TORC) Transforms. } For the TORC defense, the delay transform increases the timing delay

field of  \texttt{CacheHitEvent}'s specification so that the observable times of a cache hit and miss are the same.

For checking a security assertion (i.e., non-interference),
we carry out two transforms (see \S\ref{subsub:non_interference}). The first transform duplicates the machine state specification and event specification but the initial contents can be different. The second transform relates the initial states of the two machines. 
Initially, the presence or absence bit can be different in the two machines  but the rest of the machine state is the same in both machines.

The assertion is that 
the completion bit is set at the same time on both machines. Thus, whether the  cache line is present/absent (potentially a secret)  cannot be observed via the timing of a cache hit.

\textbf{Delaying Speculative changes to Remote Cache lines (DSRC) Transforms. } For the DSRC defense (see \S\ref{section:background}), the event transform modifies the \texttt{CacheHitEvent} to behave differently based on the value of a speculation bit. 

If the speculation bit is not set initially, then the execution sequences are the same as the baseline. 
If the speculation bit is set initially, the \texttt{CacheHitEvent}  is rejected and followed by a  \texttt{ReturnToCoreEvent},  a \texttt{ReIssueEvent}  and then another (restarted) \texttt{CacheHitEvent}.  The introduced \texttt{ReturnToCoreEvent} signals to the core that the cache hit cannot occur speculatively due to a possible coherence state change. The introduced \texttt{ReIssueEvent}  occurs once the core has verified that the load is on the right path. The \texttt{ReturnToCoreEvent} and \texttt{ReIssueEvent} prevent the \texttt{CacheHitEvent} from modifying the coherence state until the core has verified that the load is on the correct path. The rest of the execution following the restarted \texttt{CacheHitEvent}  is the same as the baseline.

The security assertion is again a form of non-interference. The first transform duplicates machine state and event specifications. The second transform relates the initial states of the two machines as follows. The  address bits of the load can be different, but the rest of the machine state is the same. The speculation bit is set to one (for both machines). The assertion is that the cache state and the completion bit are always the same on both machines.
Thus, the influence of a speculatively accessed   address (potentially a secret) cannot be observed.

\subsection{Composing the Transforms and Checking Security}
\label{sub:secure_combination_torc_dsrc}

We start step (1) of the workflow with the baseline model. We choose a defense from a selection of defenses (here, we have just two defenses, TORC and DSRC). We pick TORC first in this case and apply its transforms on the baseline model. Maestro checks whether the non-interference property is satisfied.  In this case, it finds that property is satisfied and reports the absence of a counterexample in step (3). In step (4), we check whether we have integrated all defenses from our selection and then move to step (2). In this case, we choose DSRC and compose it with the baseline + TORC model.

In step (3), the composed model is checked by Maestro and it  reports  a counterexample. This counterexample shows that the non-interference  properties of TORC are violated because the composed event specification has a double-cache-miss delay on the \texttt{CacheHitEvent} path. The long delay is observed on the machine that has the presence bit set, when the sharer bit for the load-issuing core is zero on both machines. 

The source of the long delay is an MDAV between TORC and DSRC. DSRC causes the \texttt{CacheHitEvent} to occur twice, once when it is rejected and the second time when it is restarted. TORC applies a miss penalty delay on both occasions. This causes a double delay in total, which re-introduces the timing channel.

\textbf{Revising the Defense(s).}
The counterexample demonstrates that the reject/restart behavior of the DSRC hit path causes uneven application of TORC delays.

A two-part change to resolve the timing difference is to slow down misses and to speed up hits to an equal timing.
First, we modify the miss case of DSRC so that a \texttt{CacheMissEvent}  creates a   \texttt{ReturnToCoreEvent} and a \texttt{ReIssueEvent}  on the speculative cache miss path.
Second, we modify the hit case of DSRC so that the \texttt{CacheHitEvent} on the speculative path is replaced by an introduced  \texttt{SpecCacheHitEvent}  that is not affected by TORC's transform. Hence, the speculative path has no TORC delays on a hit or a miss. The non-speculative path always has TORC delays on cache hits. This satisfies the non-interference property. 

We also considered another revision that would slow down both hit and miss paths to the same timing. In addition to being slower, it requires changes to both TORC and DSRC's transforms.
We have presented both the insecure composition and various revised defenses in the online supplement~\cite{supplement}.

\subsection{Using an Alternate Baseline Model} 
\label{sub:alternate_model}
Another possibility is to use a different baseline model  where the coherence bit is always 1 (shared), which is achieved by an additional initial state constraint added to the baseline model. This corresponds to the effect of a change from an MESI to an MSI coherence protocol. In this case, neither of the  TORC or DSRC non-interference properties are violated. There are no MDAVs because the speculative execution does not trigger  reject/restart behavior.  
We provide the full details of this alternative solution in the online supplement~\cite{supplement}.
{\color{black} \textbf{Workflow Optimization. }

 We can apply 
the two integration workflows in parallel, thus covering the required security property without any manual intervention. It is possible to extend this strategy to run multiple parallel workflows for faster coverage of security requirements. 
}

\section{Example: Scaled-Up  Model}
\label{section:attack_formulation}

In this   section and the next, we study the TORC and DSRC defenses on a simulator. This is an example of the second step of our two-step MDAV screening methodology.
 In this section, we study an implementation of an attack based on the counterexample in \S\ref{section:torc+dsrc}, on both improperly and properly combined variants of TORC and DSRC. In \S\ref{subsection:attack_experiments}, we simulate the attack in a GEM5 simulator.

 \subsection{Implementing the Attack Environment}
\label{subsection:implementation}
We build our attack environment on top of a cache coherence protocol in an inclusive three-level cache using a MESI protocol similar to GEM5's inclusive 3-level cache protocol~\cite{ruby}. We integrate TORC and DSRC support onto a two-core, out-of-order processor similar to a GEM5 O3~\cite{lowe2020gem5}, with speculation support (e.g, branch speculation and load-reordering speculation). 

 \subsubsection{TORC} 
 Suppose that one core (the transmitter)  accesses a cache line address and the line is loaded into the LLC. Then,  the other core (the receiver) accesses this \emph{remote} cache line in the LLC.
 The remote cache line hit in the LLC needs to be delayed, so a main memory access is sent to create a delay. We buffer the response from the cache hit in a private buffer near the receiver core until the delay access returns and then release it to the core~\cite{ramkrishnan2020first}.

\subsubsection{DSRC}

DSRC and similar defenses require
 hardware support to identify  instructions that can be affected by speculative attacks.

 There are two commonly used \emph{speculation protection decision models} (SPDM): BranchShadow~\cite{yan2018invisispec,yu2019speculative,aimoniotis2023recon} and ROB-Head~\cite{yan2018invisispec,ainsworth2021ghostminion,ainsworth2020muontrap}. In \emph{BranchShadow}, the load is issued speculatively when it is in a branch shadow, i.e., there is an older branch in the ROB which has not yet been resolved. In the \emph{ROB-Head} model, all loads are issued speculatively, unless the load was at the head of the ROB at the time of its issue. 
A speculation protection flag that represents the results of SPDM is added to a load 
request (by checking the ROB) before it is sent to the cache. A zero value for the flag indicates that it is not a protected load and hence, does not need to trigger DSRC feedback. A  one value indicates that it could be an unsafe speculative load and needs to be protected. It does need to trigger DSRC feedback. GETS\footnote{A GETS request in a MESI protocol refers to a read issued by a cache controller to the next cache level upon a read miss. It is to get data and to update coherence state.} requests from L1 to L2, and from L2 to L3 (i.e., LLC), also contain this flag. 

\emph{DSRC Cache Feedback.} The flag is finally used at the point that  a cache hit on an L3 cache line is checked for remote E/M coherence state (i.e., the sharer bit is not set for the receiver core and the coherence state is E or M). In case the flag is set, an additional check is carried out. If remote E/M coherence state is found, then a signal is sent back to the receiver core, as a REMOTE-EM response message. If the flag is not set, the regular MESI protocol is applied. If REMOTE-EM is returned to the receiver core, it re-issues the load when/if declared safe by the speculative protection decision model.

\subsubsection{Mitigations}
We implement more concrete versions of the two repaired models that we implemented and evaluated  using Maestro (see \S\ref{sub:secure_combination_torc_dsrc}, \S\ref{sub:alternate_model}).
First, the timing equalization  of \emph{Delay Speculative changes on Remote Miss} (\textbf{DSRM}) is implemented by sending a REMOTE-EM response message back to the private caches on both cache hit of a remote cache line and cache miss. The timing equalization  happens only if the GETS had its speculation protection flag set to one. The other aspects of the MESI protocol, such as coherence states, are unchanged. 
Second, in \emph{start-with-S MESI} (\textbf{SS-MESI}), we add a control flag to the LLC which directs all load misses in the LLC to start in the S state instead of the E state. 

\subsection{Implementing the Attack Code}
\label{subsection:attack_impl}
We propose an attack that builds on the Maestro counterexample.
A central part of the attack is a \emph{load-right-path-branch-shadow} 
 (LRBS) probe, which is a code sequence executed by the receiver. 
First, we discuss three key instructions that are a part of the 
 LRBS probe (\S\ref{subsub:key_instruction}).  Second, we  use a timing diagram to explain how the LRBS probe functions (\S\ref{subsub:timing_diagram_miss}).  
Third, we describe an \texttt{x86\_64} implementation of this probe (\S\ref{subsection:asm_description}).

\subsubsection{Key Instructions in the Attack Code}
\label{subsub:key_instruction}

First, we have the \emph{Load-Before-Branch} (LBB) instruction, which incurs a cache miss during execution. This amplifes speculation effects (i.e., it creates a large speculation window).
LBB accesses occur to a cache line address that is not touched by the transmitter core and is under the control of the probe only.
Second, we have a branch instruction, which is dependent on the result of the LBB instruction, and which later resolves as not taken. This branch  creates a speculation window within the probe. Third, we have the \emph{Load-After-Branch} (LAB) instruction,  which is on the not-taken (i.e., the right or the correct) path. This LAB instruction exploits the speculation to trigger redo operations. The LAB load's cache line address corresponds to the cache line address that is touched by the transmitter core.

\textbf{Training the Probe. }We first train the branch predictor by running the probe several times. 
The branch predictor is trained as `not taken', so that the core executes the LAB instruction speculatively before branch resolution. After training has completed, the probe is ready to detect the side-effects left behind by the transmitter.

\begin{figure}
  \includegraphics[width=\linewidth]{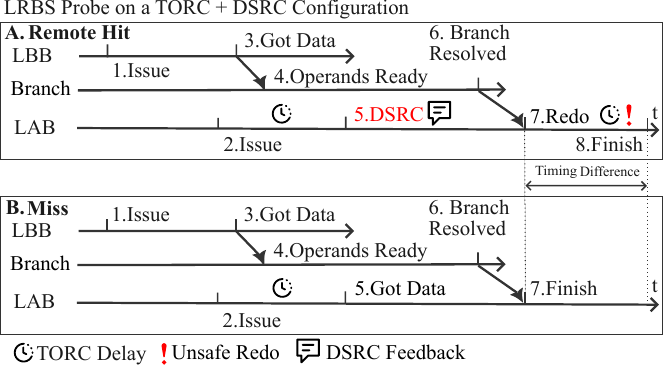}
  \caption{\textmd{A timeline showing how the MDAV caused by sending coherence information to the core is exploited using the LRBS probe on a TORC + DSRC configuration. In case A, the timeline of the probe is shown, when a remote cache line in the E state is present in the LLC (slower due to redo). In case B, the timeline is shown when there is a cache miss (faster).}}
  \label{fig:ftm_breaking_receiver}
\end{figure}

\subsubsection{Creating Timing Differences in  TORC + DSRC}
\label{subsub:timing_diagram_miss}

The probe's goal is to trigger redos conditionally according to the presence or absence of a remote cache line. Figure~\ref{fig:ftm_breaking_receiver} shows an example timing diagram which triggers redos. There are two cases, one is for a cache hit  on a remote cache line (case A) and the other is for a cache miss (case B).

\textbf{Remote Cache Line Hit.}

We first discuss the event timeline for a remote cache line hit scenario. 

The LBB and LAB instructions issue load accesses into the cache. The LBB 
gets dispatched to the execution stage first (time-point \emph{1}). Due to its earlier issue, the LBB load obtains data from the cache first (time-point \emph{3}). The branch instruction's resolution condition has a data dependency on the speculative load access (shown by an arrow from time-point \emph{3} to time-point \emph{4}).  Thus, the LBB load miss is crucial in extending the branch resolution until a later time point, as opposed to a much earlier resolution if there were an LBB cache hit. 

Meanwhile, the LAB load gets issued speculatively (time-point \emph{2}), for performance reasons, while it is still in the branch shadow. Soon after, the branch condition is available to the branch instruction (time-point \emph{4}). Sometime after that, the branch is resolved (time-point \emph{6}). The LAB load obtains a response, signalling remote E state to the core, 
after the cache has applied a TORC delay on its remote cache line hit (time-point \emph{5}). However, the LAB load cannot commit yet because it was issued speculatively into the cache and thus it did not receive any data upon a remote cache line hit. Hence, the LAB instruction waits in the pipeline until the branch is resolved (time-point \emph{6}). The core then re-issues the LAB load to the cache (time-point \emph{7}), where the load sees a TORC delay and finally returns the data to the core (time-point \emph{8}).

Time-points $5$ and $7$ are crucial parts of the exploitation of the vulnerability. At time-point $5$ the DSRC feedback occurs and at time-point $7$ the unsafe redo operation (indicated by a red exclamation mark) occurs. The DSRC feedback is abused by the attacker to trigger the redo, which ultimately causes a secret-dependent timing difference.

\textbf{Cache Miss.}
\label{subsub:timing_hit}
During a cache miss (Figure~\ref{fig:ftm_breaking_receiver}-B),  a similar event sequence occurs, except for two key differences in the LAB's timeline. The first difference is that, at time-point 5, data is returned to the core but without a coherence state. Second, there is no redo step for the LAB, so it finishes earlier.
The difference in timing measurement between a cache miss and a remote cache hit is indicated using a dotted line.

 \subsubsection{Realizing a Probe on x86\_64}
 \label{subsection:asm_description}
 Listing~\ref{lst:slow_pipe_receiver} shows an implementation of the LRBS probe. In this probe, the LBB is on line 8 and the LAB is on line 11. The LBB loads data into the \texttt{\%r12d} register. The branch instruction is on line 10, using a \texttt{jne} that is dependent on \texttt{\%r12d}. The timer measurement starts on line 5, prior to the LBB, and ends on line 14 after the branch completes its jump to line 12. \texttt{lfence}s serialize loads before (line 4) and after (line 6) the timer-start event. An \texttt{lfence}  is similarly applied to the timer-end event. Finally, line 15 calculates the timing difference and lines 16 and 17 clear the cache lines associated with the LAB and LBB. Line 18 indicates that the timing result is recorded in the \texttt{\%eax} register. Line 19 indicates the two registers that contain the addresses of LBB and LAB loads.

\begin{lstlisting} [language=C, caption={ \textmd{An implementation of the load-right-path-branch-shadow (LRBS) probe. The functionality of the code, which is to trigger secret-dependent redo operations, is explained in \S\ref{subsection:asm_description}.} },  label={lst:slow_pipe_receiver}, captionpos=b, float]
//LRBS Probe
1 asm __volatile__ (
2 " xorq %%r12, %%r12\n" //Initialize %r12 to zero
3 " mfence \n"  //Serialize older mem instructs 
4 " lfence \n"  //Serialize loads
5 " rdtsc \n"   //Start timer and store in %eax
6 " lfence \n"  //Serialize loads
7 " movl %%eax, %%esi \n" //Save timer into %esi
<@\textbf{\texttt{\textcolor{black}{8 \hspace{0.7mm}"\hspace{0.6mm}  movl (\%2), \%\%r12d~~\textbackslash n" }}}@> //LBB load
9 " testl %%r12d, %%r12d\n" //Set equals flag 
<@\textbf{\texttt{\textcolor{black}{10 \hspace{0.0mm}"\hspace{0.0mm} jne \%=f\textbackslash n"}}}@> //Branch makes speculative shadow
<@\textbf{\texttt{\textcolor{black}{11 \hspace{0.0mm}"\hspace{0.0mm} movl (\%1), \%\%eax~~\textbackslash n"  }}}@> //LAB load (redo)
12 " %=:\n" //Jump target for the jne on line 10
13 " lfence \n" //Serialize loads
14 " rdtsc \n" //End timer stored in %eax
15 " subl %%esi, %%eax \n" //<@$\Delta T$@> = %esi - %eax
16 " clflush 0(%1) \n" //Flush LAB line address 
17 " clflush 0(%2) \n" //Flush LBB line address
18 : "=a" (time) //Output C variables gets <@$\Delta T$@>
19 : "c" (LAB), "r" (LBB) //C variables
20 : "%esi", "%edx", "%r12"); //Clobbered regs
\end{lstlisting}

\section{Attack Simulations}
\label{subsection:attack_experiments}

A common GEM5 simulator configuration is used, which includes an OoO core, three levels of cache, standard interconnects and DRAM. The detailed configuration is shown in~\cite{supplement}.

We discuss the attack simulation configurations and the attack simulation results (see \S\ref{subsub:attack_result_summary}).

\subsection{Attack Simulation Configurations} 
We implement five configurations  in GEM5 for each attack simulation: $C_1$, $C_2$, $C_3$, $C_4$ and $C_5$. 
$C_1$ corresponds to an insecure cache configuration. $C_2$ corresponds to a TORC implementation. $C_3$ corresponds to a TORC + DSRC configuration. 
$C_4$ corresponds to a TORC + DSRM configuration and $C_5$ corresponds to a TORC + DSRC + SS-MESI configuration.
The  transmitter creates no cache side-effect if the secret value to be transmitted is zero. Otherwise (secret value one), it places a remote cache line into the LLC. We carried out two measurements per attack simulation, one for a secret value of zero transmitted by the transmitter, and one for a secret value of one transmitted by the transmitter (totally 10 experiments). We ran each experiment 100 times and recorded the median timing result.

\subsection{Attack Result Summary}
\label{subsub:attack_result_summary}

\textbf{On the New Attack. } Table~\ref{tab:attack_results} (for the attack results) has one row for each of the configurations $C_1$, $C_2$, $C_3$,  $C_4$ and $C_5$. The different entries in each column indicate the timing measurements made for each attack. A difference between the timing columns indicates a successful attack.
The timings for DSRM ($C_4$) and SS-MESI ($C_5$) defenses are equal indicating that the attack is mitigated. The timing results of this attack are not representative of application performance (see~\cite{supplement}). 

\begin{table}
\small
\begin{tabular} { | p{4.3cm} | p{1.4cm} | p{1.4cm} |
}

\hline
\multicolumn{3}{|c|}{Attack Simulations Using LRBS} \\
\hline
\textbf{Defense Config} &  \textbf{Secret=0}  & \textbf{Secret=1}   \\
\hline

 \multirow{1}{*}{1. Insecure ($C_1$)} &   \text{205}  &  \text{199}  \\
\hline

\multirow{1}{*}{2. TORC ($C_2$)} &  \text{205}  &   \text{205} \\

\hline

\multirow{1}{*}{3. TORC + DSRC ($C_3$)} &  \text{205}  &  \text{364}  \\
\hline

\multirow{1}{*}{4. TORC + DSRM ($C_4$)} &    \text{364}  &    \text{364}  \\

\hline

\multirow{1}{*}{5. TORC+DSRC+SS-MESI ($C_5$)} &    \text{205} &   \text{205} \\
\hline

\end{tabular}
\\[0ex]
\caption{
 \textmd{$C_1$,  $C_2$,  $C_3$, $C_4$ and $C_5$ correspond to Insecure, TORC, TORC + DSRC, TORC + DSRM and TORC + DSRC + SS-MESI, respectively. $C_3$ does not have the
attack resilience we would expect from the union of TORC and DSRC defenses. DSRM or SS-MESI restore resilience against the LRBS probe.}}
\label{tab:attack_results}
\end{table}

\textbf{Covert Channel Bitrate Estimation. } We simulate the  LRBS-based  covert channel attack on a real machine (Xeon E5-2699 v3) using a FLUSH + RELOAD  based strategy. 
Our epoch size is 1 million cycles. For error correction, we conservatively transmit each bit 16 times and take the majority on the receiver side. The error rate is 0.3\% as measured across 16 $\times$12288 = 196608  single-bit transmissions. The effective transmission rate is 6KB/s. In this experiment, 1/0 are transmitted as accessing or not accessing a particular cache line. We simulate a long cache hit time  on remote  cache lines  hits (of DSRM)  by invoking an equivalent delay loop  (see \cite{supplement}).

\textbf{On the Original Attacks. } 
We also 
simulated the original cache-hit attacks and a Spectre attack targeting coherence state, which TORC and DSRC  respectively defend against  (see \S\ref{section:background}). 
As expected, $C_4$ and $C_5$ also mitigate them. 





\arrayrulecolor{black} 
{
\color{black}
\begin{table*}
\small
\begin{tabular} { | p{3.9cm} | p{1.8cm} | p{2.2cm} | p{1.1cm} | p{1.1cm} || p{2.2cm} | p{1.2cm} | p{1.0cm} |
}

\hline
\multicolumn{8}{|c|}{Selected Defense Model Transforms and Combinations in Integra} \\
\hline
\textbf{Defense Config} &  \textbf{Protected Component}  &
\textbf{Relevant Attack Class}  &
\textbf{Integra LoC} & \textbf{Alloy LoC} &
\textbf{Integrations} &
\textbf{Alloy LoC} &
\textbf{Num MDAVs} \\
\hline

\multirow{1}{*}{1. TORC~\cite{ramkrishnan2020first,ojha2021timecache}} &  \text{Caches} & Cache Hit~\cite{giner2025cohere+} & \text{3} & \text{1022} & \text{TORC+DSRC} & \text{1480} & \textmd{1} \\
\hline

\multirow{1}{*}{2. DSRC~\cite{ainsworth2021ghostminion}} & \text{Caches} & Spectre ~\cite{kocher2019spectre}  &    \text{29}  & \text{1480} & \text{TORC+DSRM} & \text{1661} & \textmd{0} \\

\hline

\multirow{1}{*}{3. SIDO~\cite{yu2020speculative}} &    \text{Caches, Core} & Spectre ~\cite{kocher2019spectre}  & \text{29} & \text{1445} & \text{TORC+SIDD} & \text{1370} & \textmd{1} \\
\hline

\multirow{1}{*}{4. SIDD~\cite{aimoniotis2023recon}} &    \text{Caches, Core} & Spectre ~\cite{kocher2019spectre} & \text{27} & \text{1376} & \text{TORC+SIDO} & \text{1445} & \textmd{2} \\
\hline

\multirow{1}{*}{5. DoM~\cite{sakalis2019efficient}} &    \text{Caches} &  Spectre ~\cite{kocher2019spectre}  &  \text{43}  & \text{1055}  & \text{DoM+SIDD+CI} & \text{1330} & \textmd{2} \\
\hline

\multirow{1}{*}{6. Coherent Isolation~\cite{giner2022scatter,ramkrishnan2024non}} &    \text{Caches, Core} & P+P~\cite{gruss2015cache}  & \text{15} & 1560 & \text{CI+Isolation} & \text{1560} & \textmd{1} \\
\hline

\multirow{1}{*}{7. Isolation~\cite{kiriansky2018dawg}} &    \text{Caches, Core} & Spectre~\cite{kocher2019spectre}  & \text{12} & 1480 & \text{CI*+Isolation} & \text{1950} & \textmd{0} \\
\hline

\multirow{1}{*}{8. Secure DRAM Refresh~\cite{ainsworth2021ghostminion}} & \text{DRAM} &  RowHammer~\cite{mutlu2019rowhammer}  & \text{23} & 1773 & \text{SDR+Isolation} & \text{1885} & \textmd{1} \\
\hline

\end{tabular}
\\[0ex]
\caption{
{ \color{black}
 \textmd{Selected defenses,  integrations and MDAVs, modeled and checked by Maestro. Each experiment completes within 2  minutes on a 16G Macbook Air M4. Maestro finds 8 MDAVs and enables $\approx$ 15x LoC Alloy ratio. The three Maestro DSL baseline models used, each contain about 150 LoC (see \S\ref{sub:generality}).}  }}
 
\label{tab:generality}
\end{table*}
}


\section{{Generality and Scalability Evaluation}}
\label{section:miscellaneous_evaluation}
{\color{black} We discuss miscellaneous issues relating to generality and scalability. \S\ref{sub:generality} presents 8 MDAVs and two resolved integrations. \S\ref{subsection:maestro_vs_pensieve} presents runtime results for two stress tests. The Maestro framework's implementation effort is $\approx$ 9K LoC of Python. It generates more than 11K LoC for Alloy defense models from $\approx$ 450 lines of Maestro DSL and less than 200 lines of Integra DSL. Table~\ref{tab:generality} presents the key results of the MDAV investigation. 
}

{
\subsection{Demonstrating the Generality of MDAVs. } 
\label{sub:generality}
\color{black}

The key MDAVs of concern are between TORC~\cite{ojha2021timecache,ramkrishnan2019new,yan2019secdir}, DSRC~\cite{ainsworth2021ghostminion}, SIDD~\cite{aimoniotis2023recon}, SIDO~\cite{yu2020speculative}, DoM~\cite{sakalis2019efficient}, Isolation~\cite{kiriansky2018dawg}, CI~\cite{ramkrishnan2024non} and SDR~\cite{bostanci2025understanding}, as introduced in \S\ref{section:background}.

\textbf{Eight MDAVs. } 
There are five  causes for these MDAVs.
\begin{enumerate}
\item
A longer or shorter speculation window due to another defense (TORC+DSRC, TORC+SIDO).
\item
Incorrect usage of a declassification bit by a defense  (TORC+SIDD), (SIDD+CI) and (SIDD+DoM+CI).
\item
Interference with data-oblivious operations to make them secret dependent (TORC+SIDO).  
\item
Insecure coherence transactions unintentionally triggered by a defense (CI+Isolation). 
\item
Unexpected DRAM refresh delays that a defense does not account for (SDR+Isolation).
\end{enumerate}



\textbf{Two Resolved MDAVs. } TORC+DSRM and TORC+SS-MESI avoid MDAVs by adding more delays or by getting rid of the key state that caused the delay (see \S\ref{section:torc+dsrc}).
CI*+Isolation avoids MDAVs using one-way coherence operations, in scenarios where one-way information flow is permissible. 

}



{
\color{black}
\subsection{Scalability of Maestro}
\label{subsection:maestro_vs_pensieve}
An important dimension of scalability is to increase the number of bits and see how the runtime responds. We expressed 2500 bits of FIFO buffers (each 25 bits) using Maestro. The total time to solve this problem (for 10 steps) was less than 2 minutes for the SAT solver \texttt{plingeling.parallel}. The CNF generation step (Java) took less than $30$ minutes using Alloy's default settings. 
In another experiment, we set up a cycle-accurate stress test on a system resembling a 2-core OoO processor (ROB, reservation station, issue queue and load-store queues). 
It completed 20 steps in less than 2 minutes, while checking 1550 bits of state.

\textbf{High Cycle Count. } To stress Maestro's translation approach, we use two events with 62 cycles delay between them. Maestro's check completes in  under 2 seconds but it takes more than 200 seconds using a na\"{i}ve cycle-by-cycle approach. The timing difference increases with delay.

\textbf{Multiple Defenses. } To stress Maestro,  4 defenses are combined, namely, TORC, DSRC, SIDD and SIDO, which semantically compose with the same baseline model. The composed model detects an MDAV within 1 minute.

}

\section{Discussion and Related Work}
\label{section:related_work}
{\color{black} \textbf{Do MDAVs cause side-channels? }  
Maestro checks non-interference properties, which when true indicate an absence of side-channels and covert channels. Side-channels can be more dangerous because they involve unwilling participants. The new covert channel (see Table~\ref{tab:attack_results}) can be repurposed to target side-channel prone software (e.g.,  \texttt{mbedTLS}~\cite{mbedTLS}) due to secret-dependent memory access patterns (e.g., T-tables). 
Similarly, an improper integration of TORC and SIDO can result in a speculative leakage  because data-oblivious operations on the wrong path are affected by TORC (see \S\ref{section:miscellaneous_evaluation}).} 

{\color{black}
\textbf{Detecting Implementation Issues in the Second Step.}
Consider a model where  all vulnerable shared data have speculative accesses delayed based on a sharer bit in the page table.  However,  many complex changes in the system software (e.g., the Linux kernel) need to be considered to update the sharer bit accurately, due to complex page management (compare~\cite{rev_mapping} which requires new instructions). This prohibitive complexity becomes apparent in building a simulation but can be omitted in the first step.} 

\textbf{Early-Stage Modeling.}
Pensieve~\cite{yang2023pensieve} proposes a  modeling discipline the implementation of each defense is a module. However, as a consequence of RTL-like modeling, the introduction of a new module requires changes to other modules that the new module communicates with, such as wire connections.    
Compared to Maestro, Pensieve does not support event-based modeling or a composition approach where changes are confined to a single module. 

Another approach similar to Maestro's abstraction level has designers specify an event graph applied to concerns of memory consistency~\cite{wickerson2017automatically,hsiao2021synthesizing,zhang2024pipegen,lustig2016coatcheck,manerkar2018pipeproof,manerkar2017rtlcheck,manerkar2015ccicheck,trippel2017tricheck} and speculative or other leakage~\cite{ponce2022cats,mosier2022axiomatic,trippel2018checkmate}. One important difference is that these systems often lack an explicit notion of time. \textcolor{black}{The work in ~\cite{hsiao2024rtl2mmupath} supports cycle-accurate event graphs like Maestro but lacks 
transforms for composition, and multi-cycle delays between parent and child events.}

\textbf{Verification and Fuzzing. } 
Maestro is an early-stage integration complement to greater detail level analysis~\cite{rojas2021low,saravanan2024formal,lau2024specification}, including RTL-level~\cite{hsiao2024rtl2mmupath,tan2025rtl,dinesh2025scalable,chen2023hypfuzz,ceesay2024mucfi,deutschmann2024scalable,deutschmannfastpath}, cell-level~\cite{solt2022cellift} and  gate-level~\cite{zhao2024static}. 
A parallel area of research is hardware fuzzing, where the strategy is to use an  RTL input~\cite{solt2024cascade,xu2025dejavuzz,de2025phantom,borkar2024whisperfuzz} or  defense implementations on  microarchitectural simulator~\cite{fu2025amulet}. 

\textbf{Other Defenses. } There are a host of other microarchitectural defenses~\cite{kiriansky2018dawg,saileshwarbespoke,dessouky2021chunked,giner2022scatter,qureshi2018ceaser,saileshwar2021mirage,unterluggauer2022chameleon,dessouky2020hybcache,ainsworth2021ghostminion,ainsworth2020muontrap,yan2018invisispec,bahmani2021cure,lee2019keystone,pashrashid2023hidfix} for other microarchitectural attacks than those discussed so far~\cite{purnalprime+,briongos2020reload+,wan2021volcano,chen2021leaking,pessl2016drama,khaliq2021timing,tan2021invisible,gruss2016rowhammer,mutlu2019rowhammer,murdock2020plundervolt,skarlatos2021jamais,murdock2020plundervolt}. Attacks are mitigated by defensively modifying both cache~\cite{omar2020ironhide,saileshwarbespoke,dessouky2021chunked,ramkrishnan2024non,giner2022scatter,kaur2024rspp,zhao2023untangle,townley2022composable} and non-cache~\cite{loughlin2021dolma,harris2019cyclone,anagnostopoulos2018overview,khaliq2021timing,chowdhuryy2024ivleague,juffinger2025secret} components. MDAVs are left to future study.

%

\section{ Conclusion }
\label{section:conclusion}

Microarchitectural security is receiving increasing attention due to the rise of microarchitectural attacks. However, multiple defenses are not often studied together in today's literature. This has led to a lack of integrated designs that are checked to be free of microachitectural defense assumption violations (MDAVs). Therefore, we propose a two-step methodology to study defenses in composition. 
For the first step, we propose and implement a modeling framework, \emph{Maestro}, and a workflow for iterative semantic composition. 
{\color{black} Maestro reveals eight MDAVs between state-of-the-art defenses, enables compact expression (15x Alloy LoC ratio), enables seamless semantic composition and eliminates 100x performance degradations. }
For the second step, we implement an integrated state-of-the-art defense on a microarchitectural simulator.
In our evaluation, we propose a new covert channel attack that targets the MDAV and show that the defense blocks both previously known attacks and our new attack.

Overall, this work shows that it is crucial to take MDAVs into consideration when we integrate multiple defenses into a microarchitecture. Our proposed methodology can help detect and evaluate potential MDAVs at an early stage of such an 
integration.




\bibliographystyle{IEEEtranS}
\bibliography{refs}

\end{document}